\documentstyle[11pt,aas2pp4]{article}
\lefthead{de Blok \& McGaugh}
\righthead{Surface brightness and density}
\begin{document}
\title{Does Low Surface Brightness Mean Low Density?}
\author{W.J.G.~de~Blok}
\affil{Kapteyn Astronomical Institute\\ P.O.~Box 800\\9700 AV
Groningen\\The Netherlands}
\authoraddr{P.O.~Box 800, 9700 AV Groningen, The Netherlands}
\and
\author{S.S. McGaugh}
\affil{Department of Terrestrial Magnetism\\Carnegie Institution of
Washington\\5241 Broad Branch Road NW\\Washington, DC 20015}
\authoraddr{Carnegie Institution of Washington,
5241 Broad Branch Road NW, Washington, DC 20015}

 \begin{abstract} We compare the dynamical properties of two galaxies at
identical positions on the Tully-Fisher relation, but with different
surface brightnesses.  We find that the low surface brightness galaxy
UGC 128 has a higher mass-to-light ratio, and yet has lower mass
densities than the high surface brightness galaxy NGC 2403.  This is
true for the gas surface density, the stellar surface density, and the
total mass density. 

 \end{abstract} 

\section{Introduction}

Rotation curves of disk galaxies are observed to remain flat at large
radii, and do not show the Keplerian decline expected on basis of the
visible matter.  This is usually explained by dark matter (DM) that
surrounds the visible parts of these galaxies.  This DM is generally
assumed to be in the form of a massive halo, which stabilizes the
embedded visible disk and dominates the gravitational field in the outer
parts. 

Motivated by the observations that the properties of the visible
components of galaxies change along the Hubble sequence (see Roberts \&
Haynes 1994), attempts have been made to find out whether the properties
of the DM halos also change systematically with  galaxy type
or other parameters (e.g.  Tinsley 1981, Athanassoula, Bosma
\& Papaioannou 1987, Kormendy 1990, Broeils 1992).  A general conclusion
is that the importance of DM increases from early towards late-type
galaxies. 

It is interesting to know how the DM fraction depends on physical
properties like galaxy luminosity, disk size and surface brightness. In
general, it is believed that luminosity correlates with the total mass
of the halo as it should if the baryonic mass fraction and mass to light
ratio are not strongly variable.  Here, we wish to address whether the
mass {\it distribution\/} also follows that of the light.  This contains
two separate questions: {\bf 1)} does the baryonic mass surface density follow
in the obvious way from the surface brightness, and {\bf 2)} is the dark
matter distribution related to that of the light. 

The latter question can already be answered in the affirmative from
the Tully-Fisher (TF) relation (Zwaan et al.\ 1995).  Here, we confirm this
with rotation curve decompositions of two galaxies selected for their
nearly identical location in the TF plane, but with very
different surface brightnesses.  This choice is made to isolate galaxies
whose gross properties are similar, so that differences in their light
distribution have testable implications for their mass distribution. 

\section{The data}

Two galaxies which occupy identical positions on the TF relation are NGC
2403 and UGC 128.  Both have almost identical luminosities ($M_B \simeq
-19$, [$H_0 = 75$ km s$^{-1}$ Mpc$^{-1}$ throughout this paper]), and
almost identical maximum rotation velocities ($V_{\rm max} \simeq
134$~km s$^{-1}$).  They are morphologically very similar, but have a
large difference in surface brightness.  NGC 2403 has a central surface
brightness of 21.4 $B$-mag arcsec$^{-2}$, and is therefore
representative of the group of galaxies that obeys Freeman's Law
(Freeman 1970).  The central surface brightness of UGC 128 is 24.2
$B$-mag arcsec$^{-2}$.  It belongs to the group of LSB disk galaxies
that have been extensively described in van der Hulst et al.\ (1993),
McGaugh \& Bothun (1994), de
Blok, van der Hulst \& Bothun (1995) and de Blok, McGaugh \& van der
Hulst (1996).  The central surface brightness values are corrected
towards face-on assuming the disks are optically thin.  As both galaxies
have almost identical inclinations (60$\arcdeg$ for NGC 2403 and
57$\arcdeg$ for UGC 128) any systematic error in these corrections will
affect both surface brightnesses equally.  The data from NGC 2403 were
taken from Begeman (1987) and Kent (1987), those for UGC 128 from van
der Hulst et al.\ (1993).  NGC 2403 is at a distance of 3.3 Mpc, while
UGC 128 is at 60 Mpc.  Both galaxies have well-defined rotation curves;
that of NGC 2403 was derived with a beam of $45\arcsec \times
45\arcsec$, while the rotation curve of UGC 128 was derived using a
beam of $40\arcsec \times 40\arcsec$.  We will discuss possible effects
of the difference in physical resolution between the two curves later in
this paper. 

\section{Results}
\subsection{Disk-halo decomposition}

The structural properties of DM halos are usually derived by computing
the rotation curves of the visible mass components and subtracting these
from the observed curve.  After assuming a shape for the halo (e.g.  a
pseudo-isothermal sphere), a fit is made to the residuals of the
rotation curve.  This is, however, not an unambiguous procedure.  As
only the distribution of the light of the stellar component is observed,
and the mass-to-light ratio of the stellar component $\Upsilon_*$ is not
known {\it a priori}, a value for $\Upsilon_*$ has to be assumed before
the halo can be fitted.  The fitted halo parameters depend strongly on
this value.  One way to circumvent this problem is by using the
``maximum disk hypothesis'' (van Albada \& Sancisi 1986).  $\Upsilon_*$
and the rotation curve of the stellar component are scaled up to the
maximum value $\Upsilon_*^{\rm max}$ allowed by the observed total
rotation curve. 

It is still a point of discussion whether this maximum disk solution is
a physical solution (see e.g.  Kuijken \& Gilmore 1989, Athanasoula et
al.  1987, Bottema 1995).  In many cases the surface density of the
visible matter derived using other methods is found to be smaller than
that implied by the maximum disk solution.  The maximum disk solution is
in this respect a ``minimum halo'' solution: surplus rotation velocity
resulting from DM in the visible parts of galaxies is incorporated into
the rotation curve of the stellar disk by increasing its $\Upsilon_*$. 

\placefigure{Comp_2403_128}

\subsection{Results using maximum disk}

Figure \ref{Comp_2403_128} shows the observed rotation curves of HSB
galaxy NGC 2403 and LSB galaxy UGC 128, together with the rotation
curves of the individual mass components (gas, stars, DM), derived from
a maximum disk decomposition assuming an isothermal halo.  The parameters of the fit are tabulated
in the top panel of Table 1.  Though the asymptotic velocities are the
same, the rotation curve of UGC 128 rises less quickly than that of NGC
2403 (as would be expected if the mass is distributed like the light). 
At fixed global properties $L$ and $V_{\rm max}$ the {\it shape} of 
the rotation curves thus differs with surface brightness.
Note that in UGC 128 the contributions of the gas and the stars are of
equal importance in the outer parts. The gas in UGC 128 is dynamically 
more important than the gas in NGC 2403. In NGC 2403 the maximum stellar disk
dominates over the gas everywhere, despite the fact that $\Upsilon_*^{\rm max}$ is a
factor of $\sim 2$ lower than that of UGC 128.  The maximum disk masses
of the stellar component are approximately equal. The larger scale length
of the disk of UGC 128 makes the mass distribution more diffuse, so its
peak is lower and occurs at a larger radius, thus allowing large values
of $\Upsilon_*^{\rm max}$ even with  the more shallow total rotation curve. 

Taking these large values of $\Upsilon_*^{\rm max}$ at face-value 
implies either a truly high stellar mass-to-light ratio or the existence
of an additional dark component distributed like the stars.  We can
exclude the former possibility as all other evidence points towards LSB
galaxies having unevolved disks.  For an exponentially decreasing star
formation rate and a stellar population with $B-V \sim 0.7$ (typical for
a Freeman disk galaxy) population synthesis models (Larson \& Tinsley 1977)
predict a stellar mass-to-light ratio $\Upsilon_*^{B} \simeq 2.6$, while for $B-V
\sim 0.5$ (typical for LSB galaxies) $\Upsilon_*^B\simeq 0.9$.  This is
opposite to the trend derived from the maximum disk fits.  
$\Upsilon_*^{\rm max}$ in LSB galaxies is unlikely to be
representative of the stellar population. 
 
The conclusion is that DM must dominate in UGC 128 down to a small
radius.  It is therefore not ``maximum disk'' in the
sense usually applied to HSB galaxies, where the inner part of the
rotation curve can be completely explained by the luminous matter.  This
is true of LSB galaxies generally (de Blok et al.\ 1996): DM is needed
almost everywhere to explain the observations. 

\subsubsection{Beam smearing}

A natural question is whether the shallow slope of the curve of UGC 128
cannot be caused by beam-smearing effects, as the difference in physical
resolution between the two curves is a factor of $\sim
15$.  In other words, is UGC 128 a NGC 2403 in disguise? There are
several reasons why it is very unlikely that this is the case. 

We illustrate the effects of beam smearing by smoothing the rotation
curve of NGC 2403 to the same {\it physical} resolution as that of UGC
128.  This curve is shown as the light line in Fig.  1.  It is obvious
that even at this lower resolution the NGC 2403 curve is still different
from the UGC 128 curve. The parameters of the maximum disk decomposition
of this smoothed curve are also given in Table 1.

Note that this experiment is an extreme one: smoothing NGC 2403 to the
UGC 128 physical resolution has a noticeable impact on $\Upsilon_*^{\rm
max}$.  Improving the resolution of UGC 128 to that of NGC 2403 probably
will not have as big an effect: the domination of DM in LSB galaxies 
(see Sect. 3.3) produces smooth and featureless rotation curves that are
relatively insensitive to beam smearing effects (de Blok \& McGaugh
1996).  These curves do not show the characteristic tell-tale features
of a dominant disk or bulge.  This means that for halo-dominated
galaxies the measured slope cannot be much different from the true
slope. 

Note that our exercise of smoothing the NGC 2403 curve to a lower
resolution has only increased the discrepancy between the fits to the
two curves as the difference in $\Upsilon_*^{\rm max}$ between the two
galaxies has become larger. 

\subsection{Masses of the components}

The values for the total masses of galaxies one derives from rotation
curves are always lower limits.  The flatness of the rotation curves
shows that the cumulative mass keeps increasing with radius.  When
comparing masses of galaxies one thus has to measure the mass at a
consistently defined radius to take this size-effect into account. 

We derive the masses and mass-to-light ratios of both galaxies using
three different definitions for this radius and show that all
definitions result in UGC 128 being significantly more DM dominated than
NGC 2403. These definitions are as follows:
{\bf 1)} The outermost measured radius $R_{\rm max}$.
This is the closest
we can get to deriving the total mass, but as $R_{\rm max}$ is different for
both galaxies, a significant size-effect is still present. Even at $R_{\rm
max}$ one probably considerably underestimates the total mass.
{\bf 2)} A fixed number of scale lengths: in this case
the masses within 6.2 scale lengths. This is
the preferred way of measuring radii in a galaxy, as the scale
length is intrinsic to a galaxy, and the only directly observable
measure of the size.
{\bf 3)} A fixed number of kpc: in this case 20 kpc.
This would be identical to definition 2 if all galaxies had identical surface
brightnesses.

The radii were chosen to be as large as possible within the constraints
of the observations and the respective definitions.  Table 1 shows the
values of the masses of the different components (gas, stars and DM)
using the three different definitions. 
\placetable{Tab1}

The stellar masses have been computed
by scaling the measured luminosities {\it within the respective radii}
with $\Upsilon_*^{\rm max}$; the gas masses are also measured within the
respective radii by scaling the relevant fraction of the H{\sc i} mass with
1.4 to account for the helium. Finally, the dark masses have been
computed using $M_T(R) = V(R)^2 R /G$, where $R$ is the appropriate radius
according to one of the definitions, and subtracting the visible mass.

In the cases of $R<R_{\rm max}$ and $R<6.2h$ the dark component is much
more dominant in the LSB galaxy (even under the assumption of maximum
disk).  The shapes and amplitudes of the rotation curves of other LSB
galaxies (as presented in de Blok et al.\ 1996) strongly suggest that
this is true for late-type LSB disk galaxies in general. 

In the case of $R<20$ kpc the mass of the dark component in UGC 128 is
here of the same order as that of the dark component in NGC 2403.  It
should be kept in mind, though, that 20 kpc corresponds to more than 9
scale lengths in NGC 2403, that is, at 20 kpc we are seeing the
outermost observable parts of that galaxy.  In UGC 128 20 kpc
corresponds to only 2.9 scale lengths.  
This is well within the optical disk, where we are still
sampling the rising part of the rotation curve.  We are thus
measuring two physically very different regions of the two galaxies. 
This result is therefore still consistent with the conclusion that DM is
more dominant in LSB galaxies. 

We want to stress that the maximum disk case is a worst case scenario. 
If, for example, $\Upsilon_*$ in UGC 128 is decreased to the value of
$\Upsilon_*^{\rm max}$ in NGC 2403 (or a value appropriate for the
stellar population), the contribution of the disk will
naturally become less.  The dark halo fit will have to adjust itself
accordingly and put more DM into the inner disk.  UGC 128 would thus
become even more DM dominated. 

\subsubsection{Mass-to-light ratios}

This brings us to the $M/L$ ratios: these are independent of any maximum
disk assumptions, and are simply computed by dividing the mass $M_T(R)$
with the luminosity $L(R)$ (i.e.  the mass and luminosity within
radius $R$).  As the luminosity of both galaxies is by choice equal, 
differences in the mass-to-light ratios are caused by
differences in the total masses within each $R$.
In all three cases UGC 128 has the highest value of $M/L$.
 
\subsubsection{Mass-surface densities}

This leaves the question of whether LSB galaxies have a lower baryonic
mass surface density than HSB galaxies.  That is, are LSB galaxies just
normal surface density galaxies that happen to have an anomalously lower
luminous surface density, or do they have intrinsically lower surface densities. 

\placefigure{comp_surfdens}

We illustrate this in Fig.  \ref{comp_surfdens}.  For both NGC 2403 and
UGC 128 the radial surface density distributions of the gas and the
stellar components are plotted, assuming the respective maximum disk
$\Upsilon_*^{\rm max}$ values for the stellar disk.  The surface
densities shown are therefore the maximum values that can be accommodated
within the constraints of the observed rotation curves.  The total
baryonic surface density in UGC 128 is still a factor of $\sim 5$ lower
than that in NGC 2403, despite the maximum disk assumption, and despite
the fact that $\Upsilon_*^{\rm max}$ is a factor of $\sim 2$ higher in
the LSB galaxy. 

Again,
beam-smearing effects cannot cause these results. In Fig. 2 we also show
the total baryonic surface density profile as derived from the maximum
disk decomposition of the smoothed NGC 2403 curve.
Even here, at identical {\it physical} resolutions, the
baryonic surface density of UGC 128 is consistently a factor of 2 lower
than in NGC 2403.

Fig.  \ref{comp_surfdens} illustrates an extreme case.  If we relax the
maximum disk assumption, then for all plausible assumptions UGC 128 will
have the lowest baryonic mass surface density.  One could of course
assume maximum disk in UGC 128 and minimum disk in NGC 2403, but it
should be obvious that this is unrealistic. 

The rotation curve fitting procedure in most cases prefers a 
{\it minimum} disk solution for the LSB galaxies (de Blok \& McGaugh 1996), while the rotation
curves of HSB galaxies are better fitted using a {\it maximum} disk solution.
It is therefore unrealistic to lower the
inferred baryonic surface densities of the HSB galaxy by lowering
$\Upsilon_*^{\rm HSB}$, while at the same time keeping $\Upsilon_*^{\rm
LSB}$ fixed at the maximum disk value.  While maximum disk is thus a
good working hypothesis for HSB galaxies, it is not in LSB galaxies. 

We can thus conclude from Fig.  \ref{comp_surfdens} that both the gas
surface density and the stellar disk mass surface density in UGC 128 are
lower than those in NGC 2403.  From the very similar shapes and
amplitudes of the rotation curves and gas and surface brightness profiles (de
Blok et al.  1996,1995; McGaugh \& Bothun 1994), we conclude that this
is also true in general for LSB galaxies. 

A detailed quantitative conclusion about the DM distribution would
involve discussing the freedom for interplay between the various mass
components during fitting.  This is beyond the scope of this work.  The
halo parameters $\rho_0$ and $R_C$ in Table 1 are comparable, but are
extremely sensitive to the assumed values of $\Upsilon_*^{\rm max}$. 
The different geometries of the disks and the fitted values of
$\Upsilon_*$ tend to smooth out any intrinsic differences between the
two halos by attributing them mostly to differences in $\Upsilon_*^{\rm
max}$.  However, a small difference still remains (Table 1), and becomes
more pronounced at identical physical resolution.  If the
maximum disk assumption is relaxed, differences will become only more
pronounced. For comparison we also show the halo parameters derived under
the {\it minimum disk} assumption in Table 1.

A more robust statement is that the mean mass
density enclosed by the two disks (at 6.2 scale lengths) differs by an
order of magnitude: $\langle \rho \rangle = 2.3 \times 10^{-3}
M_{\sun } {\rm pc}^{-3}$ for UGC 128 and $\langle \rho \rangle =
2.5 \times 10^{-2} M_{\sun} {\rm pc}^{-3}$ for NGC 2403. 



\section{Conclusions}

We have compared UGC 128 and NGC 2403, two galaxies at identical
positions on the TF relation, but with a factor of 10 difference in
surface brightness.  Although the global properties $L$ and $V_{\rm max}$
are identical, the {\it shape} of the rotation curves differs with
central surface brightness. The mass-to-light ratio of the LSB galaxy is always
higher than that of the HSB galaxy at any given radius.  This confirms
the relation between surface brightness and mass-to-light ratio inferred
from the TF relation (Zwaan et al.\ 1995).  The surface densities of the
gas and the stars in LSB galaxy UGC 128 are lower than in the HSB galaxy
NGC 2403.  The total mass density of UGC 128 is an order of magnitude
lower than that of NGC 2403.  If all of this applies to other LSB
galaxies (as is strongly suggested by the shapes of their rotation
curves [de Blok et al.  1996]), then LSB galaxies are true low density
objects.  


\newpage

\begin{deluxetable}{lll}
\tablecaption{Comparison of UGC 128 and NGC 2403 \label{Tab1}}
\tablehead{\colhead{Name}&\colhead{UGC 128}&\colhead{NGC 2403}}
\tablewidth{94mm}
\tablecolumns{3}
\startdata \nl
$M_B$ (mag)	& --18.9 	& --19.2 \nl
$V_{\rm max}$ (km s$^{-1}$) & 131 & 136 \nl
$\mu_{0,B}$ (mag arcsec$^{-2}$)& 24.2 & 21.4\nl
$h$ (kpc) & 6.8 & 2.1 \nl
\tablevspace{6pt}
$\Upsilon_{*,B}^{\rm max}$ & 3.0 & $1.8 (0.8)$\nl
$\rho_0^{\rm halo}$ (M$_{\odot}$pc$^{-3})$ & 6.0$\times 10^{-3}$ & 10.2(19.9)$\times
10^{-3}$\nl
$R_C^{\rm halo}$ (kpc) & 9.4 & $6.6 (4.7)$ \nl
\tablevspace{6pt}
$\rho_0^{\rm halo, min\ disk}$ (M$_{\odot}$pc$^{-3})$ & 21.7$\times 10^{-3}$ & 470$\times
10^{-3}$\nl
$R_C^{\rm halo, min\ disk}$ (kpc) & 4.0 & $0.8$ \nl
\tablevspace{6pt}
$R_{\rm max}$ (kpc) & 42.3 & 19.5 \nl
$R_{\rm max}/h$ & 6.2 & 9.3\nl
\cutinhead{\bf $R<R_{\rm max}$}
$M_{\rm gas}$ ($\times 10^{10}$ M$_{\odot}$) & $1.05 $& $0.43$\nl
$M_{\rm star}^{\rm max}$ ($\times 10^{10}$ M$_{\odot}$) & $1.68 $&
$1.39 (0.61)$\nl
$M_{\rm dark}$ ($\times 10^{10}$  M$_{\odot}$) & $14.43$& $6.26 (7.04)$\nl
\tablevspace{6pt}
$\Upsilon_B^{{\rm global},R}$ & 30.6 &  10.5\nl
\cutinhead{\bf $R<6.2h$} 
$M_{\rm gas}$ ($\times 10^{10}$ M$_{\odot}$) & $1.05 $& $0.34$\nl        
$M_{\rm star}^{\rm max}$ ($\times 10^{10}$ M$_{\odot}$) & $1.68 $&$1.36 (0.60)$\nl     
$M_{\rm dark}$ ($\times 10^{10}$  M$_{\odot}$) & $14.43$& $3.90 (4.66)$\nl     
\tablevspace{6pt}
$\Upsilon_B^{{\rm global},R}$ & 30.6 &  7.4\nl
\cutinhead{$R< 20$ {\rm kpc}}
$M_{\rm gas}$ ($\times 10^{10}$ M$_{\odot}$) & $0.62 $& $0.43$\nl
$M_{\rm star}^{\rm max}$ ($\times 10^{10}$ M$_{\odot}$) & $1.23 $& $1.39 (0.61)$\nl
$M_{\rm dark}$ ($\times 10^{10}$  M$_{\odot}$) & $5.41$& $6.26 (7.04)$\nl
\tablevspace{6pt}
$\Upsilon_B^{{\rm global},R}$ & 17.7 &  10.5\nl
\enddata
\tablecomments{Parenthesized values for NGC 2403 are results obtained
from data smoothed to the physical resolution of UGC 128.}
\end{deluxetable}

\newpage
\figcaption[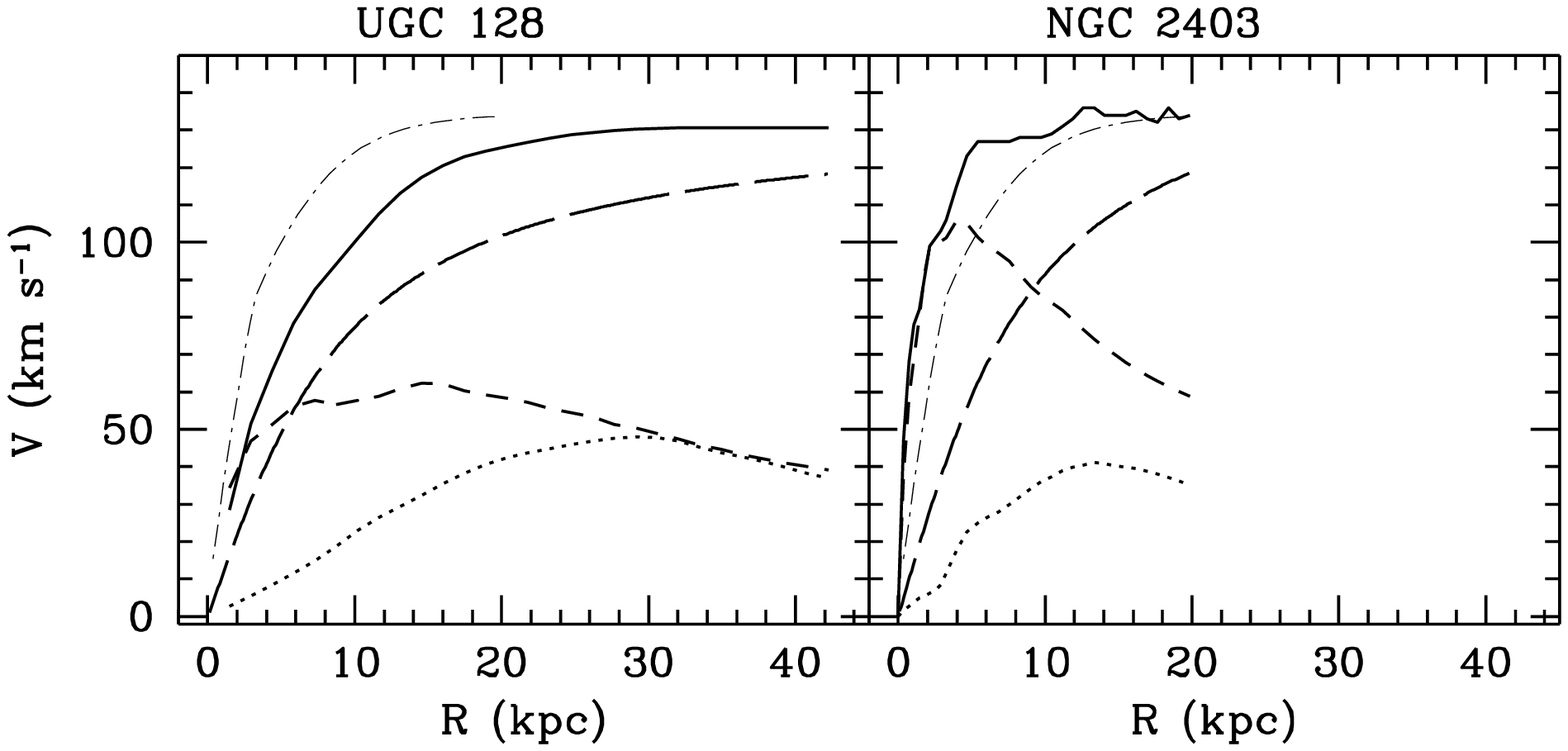]{Comparison of the rotation curves of HSB galaxy NGC 2403 and
LSB galaxy UGC 128. 
In the two panels the drawn lines are the observed rotation
curves; the dotted lines represent the rotation curves of the gas
components (observed H{\sc i} scaled by 1.4 to take He into account);
the short dashed lines are the rotation curves of the disk, scaled to
maximum disk; the long-dashed lines represent
the rotation curves of the halos under the maximum disk assumption.
The light dash-dotted lines in both panels show the rotation
curve of NGC 2403 smoothed to the same physical resolution as the UGC
128 observations. Beam smearing cannot cause us to mistake one galaxy for
the other.
\label{Comp_2403_128}}

\figcaption[comp_surfdens.ps]{Comparison of the surface densities of the
baryonic matter in UGC 128 and NGC 2403.  The stellar surface densities
have been scaled by a factor $\Upsilon_*^{\rm max}$, as inferred from
the maximum disk fits, and are therefore the maximum values both
galaxies can accommodate within the constraints of their rotation curves. 
NGC 2403 is represented by the two full lines.  The heavy top line shows
the surface density inferred from the original rotation curve, the 
starred lower line that from the rotation curve smoothed to the physical
resolution of that of UGC 128.  The baryonic surface
density in UGC 128 is significantly lower than that in NGC 2403. 
\label{comp_surfdens}}

\end{document}